\documentclass[11pt,preprint]{aastex}

\slugcomment{Accepted by the Astrophysical Journal Letters, 2005 May 16}

\shortauthors{Boboltz \& Marvel}
\shorttitle{OH\,12.8$-$0.9:  A New Water Fountain}

\begin{document}

\title{OH\,12.8$-$0.9:  A New Water-Fountain Source}


\author{David A. Boboltz}
\affil{U.S. Naval Observatory, \\
3450 Massachusetts Ave., NW, Washington, DC 20392-5420, \\
dboboltz@usno.navy.mil}

\and

\author{Kevin B. Marvel}
\affil{American Astronomical Society, \\
2000 Florida Ave., NW, Suite 400, Washington, D.C., 20009-1231, \\
marvel@aas.org}

\begin{abstract}
We present observational evidence that the OH/IR star OH\,12.8$-$0.9 is 
the fourth in a class of objects previously dubbed 
``water-fountain" sources.  Using the Very Long Baseline Array, we 
produced the first images of the H$_2$O maser emission associated 
with OH\,12.8$-$0.9.  We find that the masers are located in two compact 
regions with an angular separation of $\sim$109\,mas on the sky.  The axis of 
separation between the two maser regions is at a position angle of 
1.5$^{\circ}$ East of North with the blue-shifted ($-80.5$ to $-85.5$
\,km\,s$^{-1}$) masers located to the North and 
the red-shifted ($-32.0$ to $-35.5$\,km\,s$^{-1}$) masers to the South.  
In addition, we find that the blue- and red-shifted masers are 
distributed along arc-like structures $\sim$10--12\,mas across oriented 
roughly perpendicular to the separation axis.  The morphology exhibited
by the H$_2$O masers is suggestive of an axisymmetric wind with the 
masers tracing bow shocks formed as the wind impacts the ambient 
medium.   This bipolar jet-like structure is typical of the three other 
confirmed water-fountain sources.   When combined with the previously 
observed spectral characteristics of OH\,12.8$-$0.9, the observed 
spatio-kinematic structure of the H$_2$O masers provides strong evidence 
that OH\,12.8$-$0.9 is indeed a member of the water-fountain class.   

\end{abstract}


\section{INTRODUCTION}

As asymptotic giant branch (AGB) stars evolve toward compact objects with 
surrounding planetary nebulae (PNe), mass-loss occurs in the spherically 
symmetric circumstellar envelopes (CSEs) with low ($<$30\,km\,s$^{-1}$) 
expansion velocities.  Despite the spherical symmetry in the progenitor 
objects, PNe typically exhibit morphologies, such as axisymmetric structures, 
that are clearly not formed through normal AGB mass-loss mechanisms 
\citep{TYLENDA:03}.

Maser emission (SiO, H$_2$O and OH) from AGB stars can provide insights 
into the structure and dynamics of the CSE.  In late-type stars, the conditions 
for the formation of the various species of masers are typically met in a 
progression of zones.  The SiO masers occur relatively close to the star 
and have velocities near the radial velocity of the star.  The H$_2$O and 
OH masers are found in the circumstellar wind region at progressively 
further distances from the star.  The H$_2$O and OH spectral profiles are 
often double peaked, and the velocity range covered by the OH is 
larger than that of the H$_2$O. 

There is a small but growing number of evolved stars which have been shown 
to deviate from this prototypical AGB envelope.  These objects have been dubbed 
``water-fountain" sources due to their high-velocity H$_2$O masers.  
To date there are three confirmed water-fountain sources: 
IRAS\,16342$-$3814 \citep{STM:99, MSC:03}, IRAS\,19134+2131
\citep{IMAI:04}, and W43A \citep{IMAI:02}.  The H$_2$O masers associated
with these objects are characterized by wide velocity ranges 
$\Delta v > 100$\,km\,s$^{-1}$.  Specifically, for IRAS\,16342$-$3814, 
IRAS\,19134+2131, and W43A, $\Delta v$ is 259\,km\,s$^{-1}$, 132\,km\,s$^{-1}$ 
and 180\,km\,s$^{-1}$ respectively \citep[][and references therein]{LMM:92}.
These velocties are much greater than those of the typical OH/IR star in which 
differences between the blue- and red-shifted peaks of the OH masers 
are $\sim$20--40\,km\,s$^{-1}$ \citep{teLINTEL:89} and 
roughly 75\% of this range (15--30 \,km\,s$^{-1}$) for the H$_2$O masers 
\citep{EL:96}.   In fact, for the two sources (IRAS\,16342$-$3814 and W43A) that have 
detected OH masers, the H$_2$O masers have a larger $\Delta v$ than 
the corresponding OH.

In addition to the unusual spectral characteristics, the H$_2$O masers 
associated with the water-fountain sources appear to exhibit unique spatial 
structures as revealed by radio interferometric images.
In the first Very Long Baseline Interferometry 
(VLBI) study of the H$_2$O masers toward W43A, \cite{IMAI:02}
showed that the water masers are formed in a collimated precessing jet
with a true 3-dimensional space velocity of 145\,km\,s$^{-1}$.   Studies of 
IRAS\,16342$-$3814 \citep{MSC:03} and IRAS\,19134+2131 \citep{IMAI:04} 
have yielded similar bipolar maser distributions albeit without the precession 
observed in W43A.  The H$_2$O masers toward IRAS\,16342$-$3814 and 
W43A are also more extended than the OH masers in these sources.  
The bipolar jets traced by the H$_2$O masers, presumably along the polar 
axis of the star, may represent the onset of the axisymmetric morphologies 
that typify PNe.  The dynamical ages of the jets for IRAS\,19134+2131 and 
W43A are estimated to be $\sim$50\,yr and $\sim$40\,yr respectively 
\citep{IMAI:02, IMAI:04}.  That for IRAS\,16342-3814 is estimated to 
be $\sim$150\,yr \citep{CLAUSSEN:04}.  Thus the evolutionary stage that 
these water-fountain sources represent is likely very short, so such objects 
must be quite rare.

Maser emission from OH\,12.8$-$0.9 was first observed by \cite{BAUD:79} 
who classified it as a Type II OH/IR star because of its characteristic double-peaked
1612-MHz OH maser profile.   The star has been associated with the infrared 
source IRAS\,18139-1816 \citep{teLINTEL:89} which is $\sim$26\,arcsec away.   
The exceptional spectral characteristics of the H$_2$O masers toward 
OH\,12.8$-$0.9, were first highlighted by \cite{ESW:86}.  \cite{GOMEZ:94} used 
the Very Large Array (VLA) to show that the ``anomalous" H$_2$O and 
OH maser emission was coincident to within 1$''$, and therefore belonged 
to the same source.  \citet{GOMEZ:94} also found double-peaked profiles
for both the OH and the H$_2$O with the peaks of the OH separated by 
$\sim$23\,km\,s$^{-1}$ and H$_2$O peaks separated by nearly twice this 
amount, $\sim$42\,km\,s$^{-1}$.  
Although the velocity range for the H$_2$O is not as wide as the other 
water-fountain sources, \cite{ENGELS:02} found that the shape and 
variations in the spectra are consistent with the water-fountain class and 
are compatible with an axisymmetric wind.  Engels suggested that OH\,12.8$-$0.9 
may represent an earlier evolutionary stage of water fountain, and that 
the H$_2$O maser velocities may increase with increasing post-AGB 
wind speeds.   

Although the spectral characteristics provide a hint that OH\,12.8$-$0.9 is related
to the water-fountain sources, to verify that it is truly 
a member of this class requires high-resolution VLBI imaging of the H$_2$O 
maser emission.  We therefore observed the H$_2$O masers associated
with OH\,12.8$-$0.9 using the Very Long Baseline Array (VLBA) in order to 
map the maser distribution.   In this article, we present
the first VLBI images of the H$_2$O masers toward OH\,12.8$-$0.9
and discuss the nature of this interesting source.

\section{OBSERVATIONS AND REDUCTION \label{OBS}}

We observed the 22.2 GHz H$_2$O and both the 1612 MHz satellite-line and the 
1667 MHz main-line OH maser emission from OH\,12.8$-$0.9 ($\alpha = 18^h 16^m 49^{s}.23, \delta = -18^{\circ} 15' 01''.8$, J2000).   Observations of the 
H$_2$O masers occurred on 2004 June 21 
for 5 hrs starting at 04:43 UT (JD 2453177.7).  Observations of the OH 
masers occurred in two 5\,hr epochs on 2004 March 16 starting at 
11:04 UT (JD 2453081.0)  and 2004 July 25 starting at 02:29 UT (JD 2453211.6).
OH\,12.8$-$0.9 and three continuum calibrators (1751+096, J1832$-$2039, and 
3C\,345) were observed using the 10 stations of the VLBA.  The VLBA
is operated by the National Radio Astronomy Observatory
(NRAO).\footnote{The National Radio Astronomy Observatory is a
facility of the National Science Foundation operated under cooperative
agreement by Associated Universities, Inc.}    Reference frequencies
of 22.23508, 1.61223, and 1.66736\,GHz were used for the H$_2$O,
the OH satellite-line, and the OH main-line maser transitions respectively.  
The H$_2$O data were recorded in dual circular polarization using 
two 8-MHz (112.6\,km\,s$^{-1}$) bands centered on the local standard 
of rest (LSR) velocity of $-$58.0\,km\,s$^{-1}$.  System temperatures 
and point source sensitivities were on the order of $\sim$120\,K 
and $\sim$9\,Jy\,K$^{-1}$ respectively.  The OH data were recorded 
in dual circular polarization using four 0.5-MHz (92.7\,km\,s$^{-1}$) bands 
centered on the LSR velocity of $-$58.0\,km\,s$^{-1}$.  
System temperatures and point source sensitivities were on the order 
of $\sim$160\,K and $\sim$9.5\,Jy\,K$^{-1}$ respectively.

The data were correlated at the VLBA correlator operated by NRAO in 
Socorro, New Mexico.  Auto and cross-correlation
spectra consisting of 512 channels with channel spacings of 15.63\,kHz 
($\sim$0.22\,km\,s$^{-1}$) and 0.98\,kHz ($\sim$0.18\,km\,s$^{-1}$)
for the H$_2$O and OH respectively were produced.  
Calibration was performed in accordance with standard VLBA spectral-line 
procedures using the Astronomical Image Processing System (AIPS) 
maintained by NRAO.  The bandpass response was determined from scans
on the continuum calibrators and was used to correct the target source
data.  A preliminary inspection of the auto-correlation (AC) spectra for the 10
antennas at this point showed that the 1667 MHz main-line OH 
transition was not detected toward OH\,12.8$-$0.9. However, both 
the H$_2$O and the 1612 MHz satellite line of OH were detected in 
the AC spectra of all antennas.  The time-dependent gains of all antennas 
relative to a reference antenna were determined by fitting a template 
AC spectrum (from the reference antenna with the target source at a high 
elevation) to the AC spectra of each antenna.  The absolute flux density scale
was established by scaling these gains by the system temperature and
gain of the reference antenna.  Errors in the gain and pointing of the
reference antenna and the atmospheric opacity contribute to the error
in the absolute amplitude calibration, which is accurate to about
15--20\%.

 To correct any instrumental delay, a fringe fit was performed on the 
 continuum calibrator scans and residual group delays for each 
antenna were determined.  Residual fringe-rates were 
obtained by fringe-fitting a strong reference feature in the spectrum
of each maser transition.  
At this stage in the reduction, no fringes for the 1612 MHz OH 
line were detected on baselines other than 
the shortest (Los Alamos--Pie Town) VLBA baseline.  The likely
cause for this non-detection is localized interstellar scattering 
along the line-of-sight to OH\,12.8$-$0.9, which is relatively close to 
the galactic center, a direction known to exhibit anomalously
high scattering on certain lines of sight \citep{Backer:78}.
Thus, the remaining discussion applies only to the H$_2$O maser 
data.  The fringe-rate solutions from the fringe fit were applied to all
channels in the H$_2$O maser spectrum.  An iterative self-calibration 
and imaging procedure was then performed on the reference 
channel.  The resulting residual phase and amplitude corrections 
were applied to all channels in the band.

A low-resolution cube of images $512\times 512$ pixels 
($\sim$$500\times 500$\,mas) was formed covering the 
observed velocity range to locate regions of emission using a 
synthesized beam of $3.0\times 2.9$\,mas.  Two regions of emission 
were identified corresponding to spectral channel ranges from 
$-$37.8\,km\,s$^{-1}$ to $-$29.4\,km\,s$^{-1}$ and from $-$86.2\,km\,s$^{-1}$ 
to $-$75.7 \,km\,s$^{-1}$.   Full-resolution cubes of images 
$1024\times 1024$ pixels ($\sim$$80\times 80$\,mas) were generated 
for each range of channels using a synthesized beam of 
$1.98\times 0.92$\,mas.   The RMS off-source noise 
($\sigma_{\rm RMS}$) was typically 3\,mJy\,beam$^{-1}$ in all channels.

In order to extract maser component parameters, two-dimensional
Gaussian functions were fit to the emission in the individual channel
maps.  Image quality was assessed using the off-source RMS noise and the 
deepest negative pixel in the image.   A cutoff flux density was conservatively 
set at 10$\sigma_{\rm RMS}$.  Features with 
flux densities greater than this cutoff were fit with Gaussians to 
determine component parameters.  Errors in the fitted positions of identified 
features were computed following the methods outlined in \citet{CONDON:97}.
These errors ranged from 5$\mu$as for features with high signal-to-noise, 
to 60~$\mu$as for features with lower signal-to-noise.  

Because the emission from each maser feature extends over multiple spectral 
channels, it is desirable to determine a single velocity, flux density and position 
for each maser feature.  In lieu of a full three-dimensional (3-D) Gaussian fit 
to the image cube, we used a flux-density-squared weighted averaging scheme 
to calculate nominal positions and velocities for each maser from the 
previously identified features.  This method takes into account the 
increased significance of stronger emission channels.  The 
velocities and relative offset positions thus represent weighted averages 
over the number of spanned channels.  The assigned flux density is simply 
the peak value over the same range of channels.  

\section{RESULTS AND DISCUSSION}

Figure~\ref{SPOT_MAP} shows the spectral (upper sub-panels) and spatial 
(lower sub-panels) distributions of the H$_2$O masers toward 
OH\,12.8$-$0.9 from the analysis of our VLBA images.
Panel (a) shows the entire range of H$_2$O maser emission from the star.  
Panels (b) and (c) show enlarged views of the blue-shifted masers to 
the North and the red-shifted masers to the South respectively.   

Focusing on the upper sub-panel of Figure~\ref{SPOT_MAP}(a), we see 
the previously observed spectral characteristics of the OH\,12.8$-$0.9.
The 1612 MHz OH masers form a double-peaked profile with peaks at
$-$68.0 and $-$43.7\,km\,s$^{-1}$.  The center of the profile is roughly 
$-$55.8\,km\,s$^{-1}$, which is consistent with the stellar radial velocity of
$-55.5\pm 0.5$\,km\,s$^{-1}$ previously determined from the OH emission
\citep{BAUD:79,DT:91}.  The H$_2$O spectrum also has a 
double-peaked profile, however,  the H$_2$O masers have a greater 
velocity extent with peaks at $-$81.7 and $-$33.3\,km\,s$^{-1}$.  The midpoint
between these peaks ($-$57.5\,km\,s$^{-1}$) is slightly offset from the 
stellar velocity.

The lower sub-panel of Figure~\ref{SPOT_MAP}(a) shows
that the masers occupy two distinct regions oriented roughly North--South 
on the sky with an angular separation of $\sim$109\,mas.   The position 
angle of the axis of separation between the centers of the northern and 
southern maser regions is 1.5$^{\circ}$ East of North.  This double-lobed
distribution is consistent with VLBI observations of other water-fountain 
sources that show the masers occupy widely separated clumps along 
a presumed polar axis.  Separations of
$\sim$3000\,mas (6000 AU), $\sim$150\,mas (2400 AU) and 
$\sim$700\,mas (1700 AU) were 
determined for IRAS\,16342$-$3814, IRAS\,19134+2131 
and W43A respectively \citep{MSC:03, IMAI:02, IMAI:04} 
(note that the distances to all sources are somewhat uncertain, with W43A 
the most reliably known).  The linear separation of the two maser regions 
toward OH\,12.8$-$0.9 is 
approximately $109\,{\rm AU} \times D / (1\,{\rm kpc})$,
where $D$ is the distance in kpc.  Unfortunately, 
the distance to OH\,12.8$-$0.9 is unknown.   If we assume the star is 
associated with the galactic center \citep{BAUD:85}, at $D \approx 8$\,kpc, 
then the linear separation between the blue- and red-shifted masers is
$\sim$870\,AU.  

Upon closer inspection of the blue- and red-shifted maser regions 
(Figure~\ref{SPOT_MAP} (b) and (c)), we find that the masers are aligned 
along arc-like structures.  The blue- and red-shifted arcs 
are approximately 10 and 12\,mas (80 and 96\,AU at 8\,kpc)
across respectively, corresponding to an opening angle for the jet
that likely drives their formation of 10--13$^{\circ}$.  These arcs have
an orientation that is  roughly perpendicular to the separation (polar) axis.
One might expect such a morphology in the case of an edge-on 
shock front propagating outward into a slower ambient medium.  Such 
arc-like structures have also been observed for the H$_2$O masers
toward IRAS 16342$-$3814, and have been shown to have a global 
outward motion \citep{MSC:03, CLAUSSEN:04}.  Further observations of
OH\,12.8$-$0.9 will be required to measure the proper motions of its 
H$_2$O masers.

Although the radial velocity range for the H$_2$O is not as wide as the more 
well-known water-fountain sources, \cite{ENGELS:02}
found that the shape and variations in the H$_2$O maser spectra 
are consistent with this class.  The velocity difference between 
the blue- and red-shifted H$_2$O masers also appears to be increasing over time
from early values of  $\Delta v \approx 39$\,km\,s$^{-1}$ \citep{ESW:86}
to our most recent measurement of $\Delta v = 48.4$\,km\,s$^{-1}$.  
Values of $\Delta v$ based on the most dominant features in the H$_2$O 
spectra as published in the literature are listed in Table~\ref{ACCEL_TABLE} 
and plotted in Figure~\ref{ACCEL_PLOT}.  All values prior to those
in this work are determined from low spatial resolution (i.e. single dish and VLA) 
observations and are therefore subject to spatial blending of multiple maser 
features in a single spectral channel.  From Figure~\ref{ACCEL_PLOT} 
we see that $\Delta v$ is obviously not constant over time, 
and that the masers are undergoing some form of acceleration (possibly 
non-linear).   Since the errors in $\Delta v$ as determined from the literature
are not well known, we chose to perform a simple least-squares fit to the 
points assuming constant acceleration.   This fit yields a value of $
2.2 \pm 0.2 \times 10^{-8}$\,km\,s$^{-2}$ or $0.68 \pm 0.06$\,km\,s$^{-1}$\,yr$^{-1}$ 
for the relative acceleration of the blue- and red-shifted masers in 
the radial direction.  The 3-D acceleration is likely larger by a factor 
of $1/(\cos i)$, where $i$ is the inclination of the jet.  
The true 3-D velocities and accelerations of the masers remain to be 
determined through future VLBI proper motion studies. 

As mentioned previously, the distance to OH\,12.8$-$0.9 is unknown, 
which makes estimating the dynamical age of the outflow difficult.  
If we make several assumptions, namely: that the motions of 
the masers in the plane of the sky are comparable to the line-of-sight
acceleration, that the dynamical center of the outflow is the 
midpoint along the axis of separation between the two maser
regions, that the distance to the source is 8\,kpc, and that the 
masers have zero initial velocity, then an upper limit to the dynamical 
age of the outflow can be computed.   Under these assumptions, we 
compute a dynamical age of $\sim$109\,yr using the current spatial extent 
of the masers and the previously determined acceleration.  Similarly, 
a dynamical age of $\sim$71\,yr is determined using the current 
expansion velocity of 24.2\,km\,s$^{-1}$ and the above acceleration.
These two ages are roughly consistent with each other and with the ages 
computed for the three other water-fountain sources.  For comparison, 
the dynamical ages of IRAS\,16342$-$3814, IRAS\,19134+2131 
and W43A have been estimated to be $\sim$150\,yr \citep{CLAUSSEN:04},
$\sim$50\,yr \citep{IMAI:04} and $\sim$40\,yr \citep{IMAI:02} respectively.
It is interesting to note that while the small velocity spread between the 
blue- and red-shifted masers of OH\,12.8$-$0.9 relative to the other 
water-fountain sources would seem to indicate that the outflow is 
relatively young, the upper limit on the dynamical age is intermediate
to those of the other sources in the class.  Assuming the acceleration 
remains constant at 0.68\,km\,s$^{-1}$\,yr$^{-1}$, then the time to reach 
an outflow velocity comparable to the other water fountain sources
($\Delta v \approx 150$\,km\,s$^{-1}$) is only about 220\,yr or roughly twice the 
current dynamical age of OH\,12.8$-$0.9.  

\section{CONCLUSIONS}

Using the VLBA we have produced the first high-resolution interferometric 
images of the H$_2$O masers toward the OH/IR star OH\,12.8$-$0.9.  
These images show that the masers lie in a double-lobed structure 
typical of objects belonging to the water-fountain class of sources.   
We also find that the two groupings of masers are arranged in arc-like 
structures suggestive of a bow-shock morphology.  The spatio-kinematic 
structure determined from our VLBA observations provides strong 
evidence that OH\,12.8$-$0.9 is indeed a water fountain source, 
and that the outflow is accelerating in the line-of-sight direction.

In future observations, we hope to map the scatter-broadened OH masers 
using a shorter-baseline connected element interferometer such as the 
VLA or the Multi-Element Radio Linked Interferometer (MERLIN).  Such 
observations should verify that the H$_2$O masers lie further from the 
underlying star than the OH masers; a characteristic evident in the other
water-fountain sources.  We also plan to undertake a multi-epoch 
VLBI campaign to study the proper motions of the H$_2$O masers.  
Such observations should allow us to characterize the three-dimensional
motion of the masers, determine the distance to OH\,12.8$-$0.9, and 
determine a dynamical center and age for the outflow.  

\acknowledgements
KBM wishes to thank the NRAO for supporting him during a short research leave
from his day job in Washington, DC and TEK for her continued support.

%
\begin{deluxetable}{cccl}
\tablewidth{0pt}
\tabletypesize{\footnotesize}
\tablecaption{
Velocity separation of dominant water maser components.\label{ACCEL_TABLE}
}
\tablehead{
\colhead{Obs. Date} & \colhead{$\Delta\,v$} & \colhead{Spectral res.} & \colhead{Notes} \\
 \colhead{(yr)} & \colhead{(km\,s$^{-1}$)} & \colhead{(km\,s$^{-1}$) } & \colhead{Reference, features used}   \\
} 
\startdata
1985.30 & 34.7 &  0.2 & (1), strongest features - matched to hi-res ep. \\
1985.58 & 36.1 & 0.2 &  (1), strongest features - matched to hi-res ep. \\
1987.45 & 38.4 & 0.16 & (2), feature F - feature N  \\
1988.51 & 38.9 & 0.16 & (2), feature F - feature N  \\
1992.47 & 42.3 & 2.6 &  (3), (blend of blue and red 
features) \\
1995.19 & 44.5 & 0.16 & (2), feature B - feature O \\
1999.09 & 45.5 & 0.16 & (2), feature B - feature O  \\
2004.57 & 48.4 & 0.2 & (4), strongest features \\

\enddata
\tablecomments{References: (1) \cite{ESW:86}, (2) \cite{ENGELS:02}, 
(3) \cite{GOMEZ:94}, (4) this work. }

\end{deluxetable}

\clearpage

\centering{Figure Captions}

\figcaption{The H$_2$O maser emission toward OH\,12.8--0.9.  In 
panels (a), (b) and (c), the top sub-panels show the spectra formed 
by plotting maser component flux density versus LSR velocity, color-coded 
according to maser velocity.    A dark solid line in each upper sub-panel 
represents the scalar-averaged cross-power spectrum on the 
Los Alamos--Pie Town VLBA baseline.   A dashed line in 
the top sub-panel of (a) represents the scalar-averaged cross-power spectrum of 
the OH masers (with the flux density scaled by a factor of 0.5) from the 
Los Alamos--Pie Town baseline.  The lower sub-panels in (a), 
(b) and (c) plot the spatial distribution of the H$_2$O masers with point-color 
representing the corresponding velocity bin in the spectrum and 
point-size proportional to the logarithm of the maser flux density.   
Panel (a) shows all H$_2$O maser components from our VLBA observations.  
A dashed line represents the axis of separation between the centers of the 
blue- and red-shifted masers at a position angle of 1.5$^{\circ}$ East of North.  
The ``$\times$'' symbol represents the midpoint of the bipolar distribution.  
Panels (b) and (c) show expanded views of the blue- and red-shifted maser 
features respectively.  Errors in the positions of the features are smaller 
than the data points for all panels.  \label{SPOT_MAP}}

\figcaption{Plot showing the measured velocity separation ($\Delta v$) 
of the dominant water maser features as a function of time from values 
listed in Table \ref{ACCEL_TABLE}.   The line represents a  
linear least-squares fit to the data and indicates a constant acceleration 
of 0.68\,km\,s$^{-1}$\,yr$^{-1}$. \label{ACCEL_PLOT}}
\clearpage
\epsscale{1.0}
\begin{figure}[hbt]
\plotone{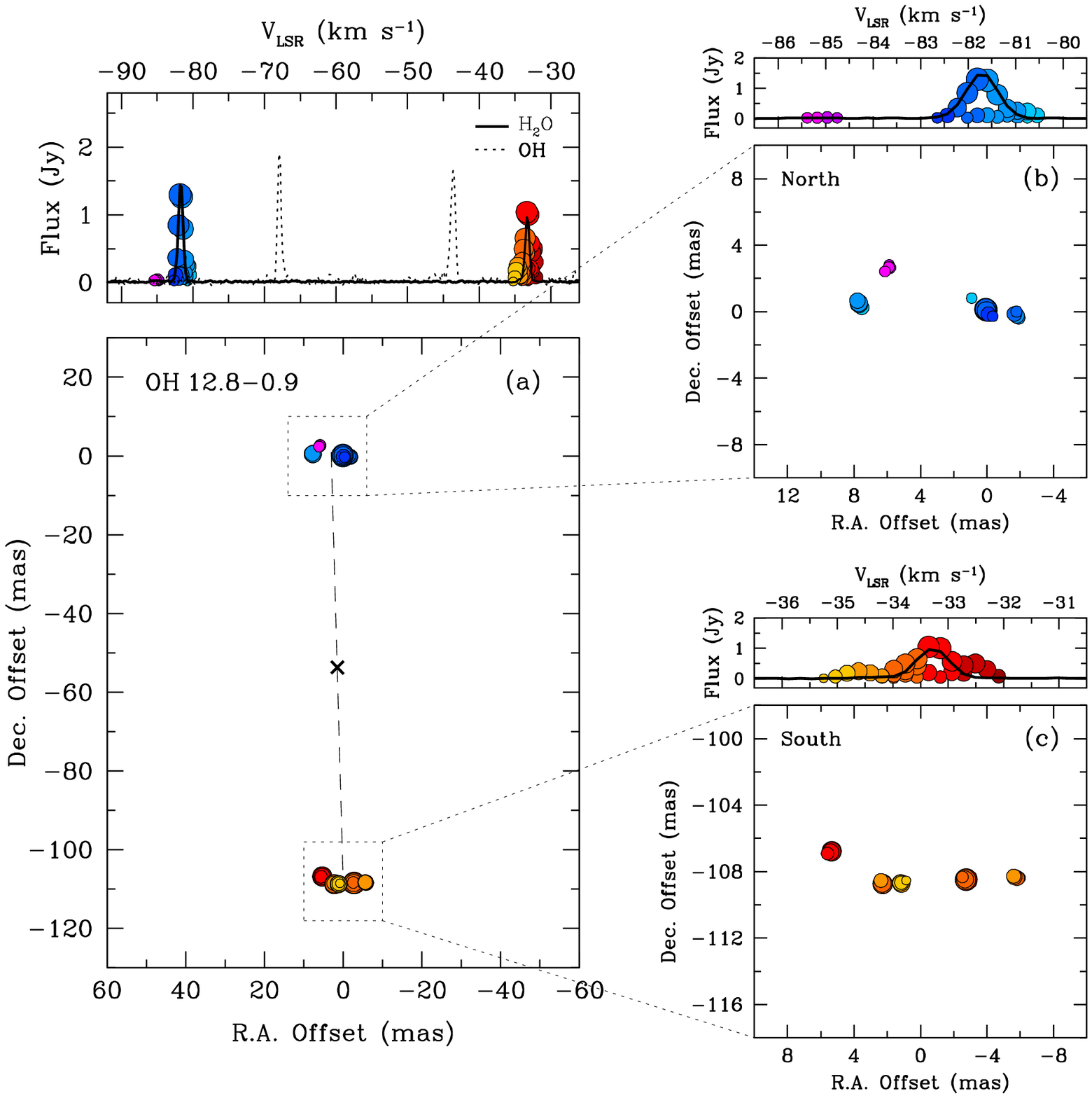}
\centerline{Figure \ref{SPOT_MAP}}
\end{figure}
\epsscale{0.75}
\begin{figure}[hbt]
\plotone{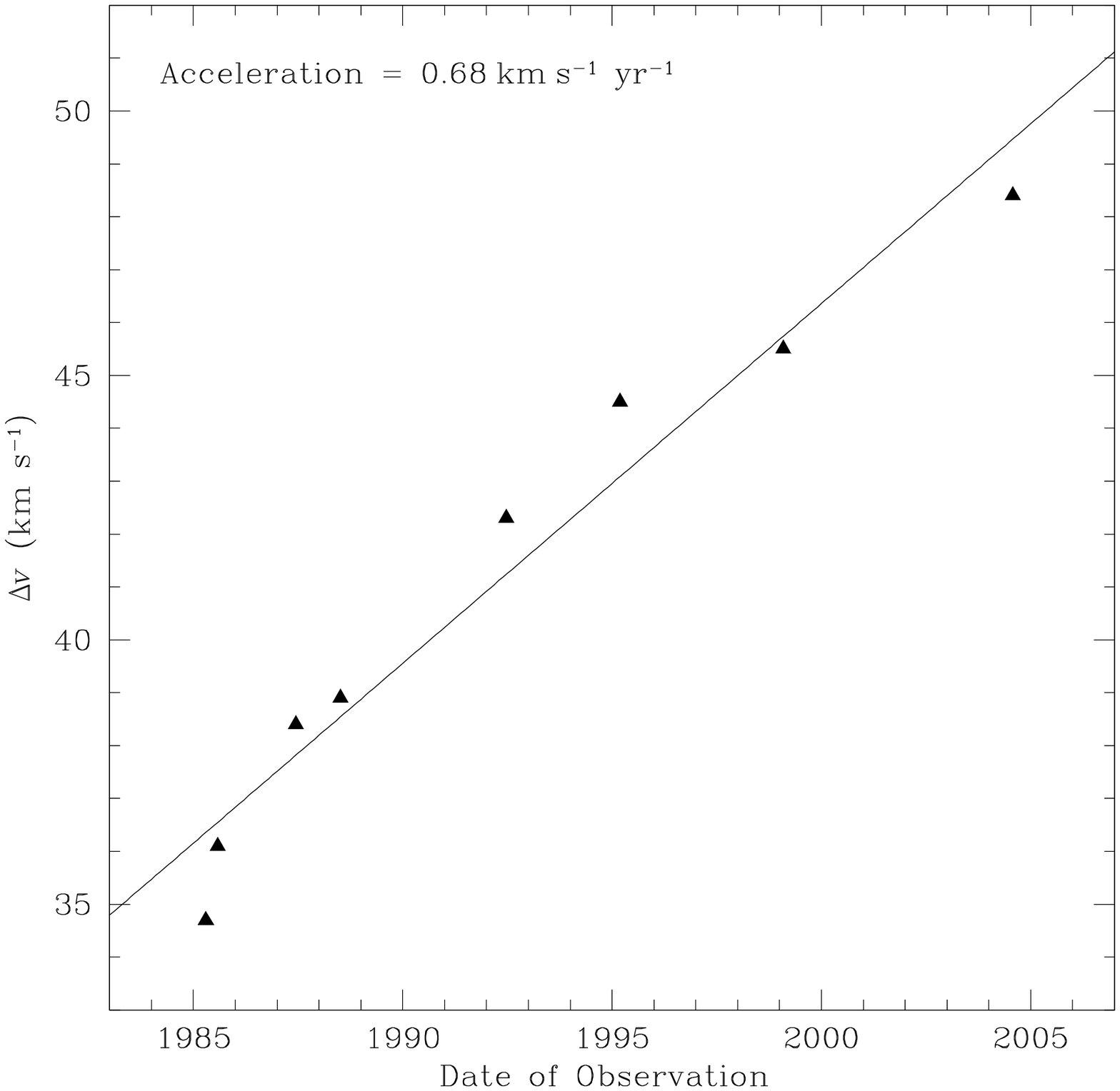}
\centerline{Figure \ref{ACCEL_PLOT}}
\end{figure}

\end{document}